\documentclass[a4paper,11pt]{article}

\usepackage{jheppub}

\usepackage{amsfonts,graphics,epsfig,subfigure}

\title{Heat engine in the three-dimensional spacetime}

\author[a,b,1]{Jie-Xiong Mo,\note{Corresponding author}}
\author[b]{Feng Liang,}
\author[a,b]{Gu-Qiang Li}

 \affiliation[a]{Institute of Theoretical Physics, Lingnan Normal University, Zhanjiang, 524048, Guangdong, China}
\affiliation[b]{Department of Physics, Lingnan Normal University, Zhanjiang, 524048, Guangdong, China}

\emailAdd{mojiexiong@gmail.com}
\emailAdd{88946654@qq.com}
\emailAdd{zsgqli@hotmail.com}

\abstract{We define a kind of heat engine via three-dimensional charged BTZ black holes. This case is quite subtle and needs to be more careful. The heat flow along the isochores does not equal to zero since the specific heat $C_V\neq0$ and this point completely differs from the cases discussed before whose isochores and adiabats are identical. So one cannot simply apply the paradigm in the former literatures. However, if one introduces a new thermodynamic parameter associated with the renormalization length scale, the above problem can be solved. We obtain the analytical efficiency expression of the three-dimensional charged BTZ black hole heat engine for two different schemes. Moreover, we double check with the exact formula. Our result presents the first specific example for the sound correctness of the exact efficiency formula. We argue that the three-dimensional charged BTZ black hole can be viewed as a toy model for further investigation of holographic heat engine. Furthermore, we compare our result with that of the Carnot cycle and extend the former result to three-dimensional spacetime. In this sense, the result in this paper would be complementary to those obtained in four-dimensional spacetime or ever higher. Last but not the least, the heat engine efficiency discussed in this paper may serve as a criterion to discriminate the two thermodynamic approaches introduced in Ref.~\cite{Frassino2} and our result seems to support the approach which introduces a new thermodynamic parameter $R=r_0$.}

\begin{document}
\maketitle
\flushbottom

\section{Introduction}
      Viewing the cosmological constant $\Lambda$ as a variable \cite{Caldarelli}-\cite{Cvetic} and identifying it as the thermodynamic pressure, the extended phase space thermodynamics \cite{Dolan2} of black holes have attracted extensive attention in recent years. Probing black hole thermodynamics in the extended phase space is of great physical significance \cite{Gunasekaran}. Firstly, one can take into consideration more fundamental theories which admit the variation of physical constants. Secondly, the Smarr relation is in accordance with the first law of thermodynamics under this frame. Thirdly, the mass of the black hole can be identified as enthalpy rather than internal energy \cite{Kastor}. In the extended phase space, not only the analogy between black holes and van der Waals liquid-gas system has been enhanced \cite{Kubiznak}, but also novel phenomena such as reentrant phase transition \cite{Gunasekaran, Frassino, Hennigar} and triple point \cite{Frassino, Altamirano2, Shaowen1} have been reported for black holes. For more details, one can read the most recent review \cite{Kubiznak2} and references therein.

    With both the thermodynamic pressure and volume defined in the extended phase space, Johnson creatively introduced the traditional heat engine into black hole thermodynamics \cite{Johnson1}. This is a natural but rather amazing proposal since the heat engines allow us to extract useful mechanical work from heat energy. Moreover, such heat engines may have interesting holographic implications because the engine cycle represents a journal through a family of holographically dual large $N$ field theories \cite{Johnson1}. It was argued that changing $\Lambda$ involves not only the change of the $N$ of the dual theory \cite{Johnson1} but also the change of the size of the space on which the field theories live \cite{Karch}. The pioneering work of Johnson was soon generalized to Gauss-Bonnet black holes \cite{Johnson2}, Born-Infeld black holes \cite{Johnson3}, Kerr AdS black holes \cite{Sadeghi1}, higher-dimensional black holes \cite{Belhaj} and other interesting aspects \cite{Setare, Sadeghi2, Bhamidipati, Caceres, Johnson4, weishaowen}.

    In this paper, we would like to extend the heat engine research to lower-dimensional spacetime, which, to the best of our knowledge, has not been covered in literature yet. Specifically, we will use the charged BTZ black holes \cite{Zanelli, Zanelli2} to construct our heat engine. It seems to be a trivial task at first glance since it was reported that charged BTZ black holes do not exhibit any critical behavior \cite{Gunasekaran}. However, we will show that as for heat engine, this spacetime is quite subtle and needs to be more careful or else one will easily make mistakes. Moreover, exploring the properties of heat engine in lower-dimensional spacetime is of strong motivations. On the one hand, lower-dimensional theories of gravity have gained renewed interest since an effective two-dimensional Planck regime is supported by many evidence. It was suggested recently the physics of quantum black holes may be effectively lower-dimensional \cite{carr}. Owing to the fact that it can be formulated as a Chern-Simon theory, 3D gravity has become paradigmatic for understanding general features of gravity. For lower-dimensional black holes in the extended phase space, attention has been focused on their critical behavior \cite{Gunasekaran}, the connections between black hole thermodynamics and chemistry \cite{Frassino2}, two-dimensional dilaton gravity \cite{Grumiller} respectively. On the other hand, the charged BTZ black hole may serve as a toy model to investigate the heat engine. It is hoped that one can obtain analytic expression of the heat engine efficiency of the charged BTZ black hole, thus providing us the first example to examine the exact efficiency formula proposed in Ref. \cite{Johnson4}. Furthermore, the result in this paper would be complementary to those obtained in four-dimensional spacetime \cite{Johnson1} or ever higher \cite{Belhaj}.

    The organization of this paper is as follows. Review of the thermodynamics of charged BTZ black holes will be presented in Sec.\ref{Sec2} . Then we will view the three-dimensional charged BTZ black hole as heat engine and investigate its efficiency in Sec.\ref{Sec3}. In the end, discussions will be presented in Sec.\ref {Sec4} and a brief conclusion will be drawn in Sec.\ref {Sec5}.

\section{A brief review of the thermodynamics of charged BTZ black holes}

\label{Sec2}
The charged BTZ black hole solution and the gauge field read \cite{Zanelli2, Gunasekaran, Frassino2}
\begin{eqnarray}
ds^2&=&-fdt^2+\frac{dr^2}{f}+r^2d\varphi^2, \nonumber
\\
F&=&dA, \;\;\;\; A=-Q\log\left(\frac{r}{l}\right)dt,
\label{1}
\end{eqnarray}%
where
\begin{equation}
f=-2m-\frac{Q^2}{2}\log\left(\frac{r}{l}\right)+\frac{r^2}{l^2}.\label{2}
\end{equation}%
The Hawking temperature and the entropy have been obtained as \cite{Frassino2}
\begin{eqnarray}
T&=&\frac{f'(r_+)}{4\pi}=\frac{r_+}{2\pi l^2}-\frac{Q^2}{8\pi r_+},\label{3} \\
S&=&\frac{1}{2} \pi r_+. \label{4}
\end{eqnarray}%
Note that $r_+$ is the horizon radius which can be determined by the largest root of the equation $f(r_+)=0$.

Identifying the cosmological constant as the thermodynamic pressure through the definition $P=-\frac{\Lambda}{8\pi}=\frac{1}{8\pi l^2}$, Ref.~\cite{Gunasekaran} derived the equation of state as
\begin{equation}
P=\frac{T}{v}+\frac{Q^2}{2\pi v^2},\label{5}
\end{equation}%
where $v=4r_+$. Based on the equation of state, Ref.~\cite{Gunasekaran} argued that the charged BTZ black holes do not exhibit any critical behavior.

The computation of the mass is quite problematic due to the asymptotic structure of the black hole solution. Ref.~\cite{Cadoni} obtained a renormalized black hole mass $M_0(r_0)$ by enclosing the system in a circle of radius $r_0$ and taking the limit $r_0\rightarrow \infty$ whilst keeping the ratio $r/r_0=1$. This mass can be interpreted as the total energy inside the circle of radius $r_0$.

An alternative approach to determine the mass is to utilize the Komar formula~\cite{Kastor}. Ref.~\cite{Frassino2} showed the first law $dM=TdS+VdP+\Phi dQ$ holds provided that the relevant quantities are defined as
\begin{eqnarray}
V&=&\left(\frac{\partial M}{\partial P}\right)_{S,Q}=\pi r_+^2-\frac{1}{4}Q^2\pi l^2,\label{6}
\\
\Phi&=&\left(\frac{\partial M}{\partial Q}\right)_{S,P}=-\frac{1}{8}Q\log\left(\frac{r_+}{l}\right), \label{7}
\end{eqnarray}%
where the mass
\begin{equation}
M=\frac{m}{4}=\frac{r_+^2}{8l^2}-\frac{Q^2}{16}\log\left(\frac{r_+}{l}\right).\label{8}
\end{equation}%
However, the Reverse Isoperimetric Inequality is violated for all $Q\neq0$ and charged BTZ black holes are always superentropic~\cite{Frassino2}.

To make charged BTZ black holes satisfy the Reverse Isoperimetric Inequality, Ref.~\cite{Frassino2} introduced a new thermodynamic parameter $R=r_0$ associated with the renormalization length scale and a new work term in the first law which is interpreted that a change in the renormalization scale leads to a change in the renormalized mass. Through this treatment, the first law and relevant quantities are changed into
\begin{eqnarray}
dM&=&TdS+VdP+\Phi dQ+KdR,\label{9}
\\
M&=&\frac{m_0}{4}=\frac{r_+^2}{8l^2}-\frac{Q^2}{16}\log\left(\frac{r_+}{R}\right),\label{10}
\\
V&=&\left(\frac{\partial M}{\partial P}\right)_{S,Q,R}=\pi r_+^2,\label{11}
\\
\Phi&=&\left(\frac{\partial M}{\partial Q}\right)_{S,P,R}=-\frac{1}{8}Q\log\left(\frac{r_+}{R}\right),\label{12}
\\
K&=&\left(\frac{\partial M}{\partial R}\right)_{S,Q,P}=-\frac{Q^2}{16R}.\label{13}
\end{eqnarray}%
Note that the above treatment retained the standard definition of the thermodynamic volume~\cite{Frassino2}.

\section{Three-dimensional charged BTZ black holes as heat engine}
\label{Sec3}
\begin{figure}
\centerline{\subfigure[]{\label{1a}
\includegraphics[width=8cm,height=6cm]{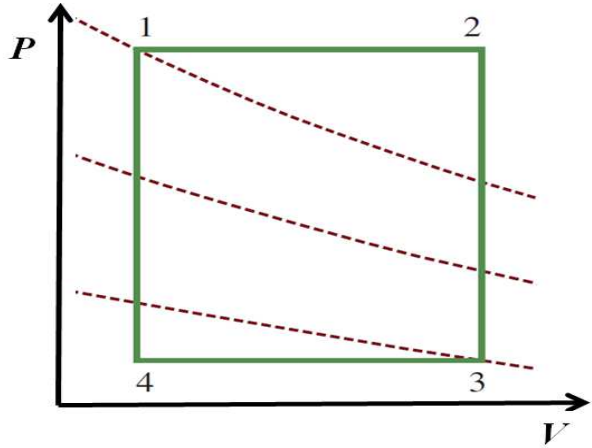}}}
 \caption{The heat engine cycle considered in this paper}
\label{fg1}
\end{figure}

In this section, we would like to define a new kind of heat engine via three-dimensional charged BTZ black holes. Specifically, we will consider a rectangle cycle in the $P-V$ plane just as done in former literatures \cite{Johnson1, Johnson2, Johnson3, Johnson4}. The rectangle consists of two isobars and two isochores as shown in Fig.\ref{fg1}, where $1, 2, 3, 4$ denote the four corners of the cycle. In this paper, we use the subscripts $1, 2, 3, 4$ to denote the relevant quantities evaluated at the four corners respectively. Below we will investigate the efficiency of the heat engine from two different perspectives.

On the one hand, if we insist that $dM=TdS+VdP+\Phi dQ$ and define the relevant thermodynamic volume as Eq. (\ref{6}), we can derive the relation among $S$, $T$, $V$ as
 \begin{equation}
T=\frac{\pi Q^2 V}{16S(4S^2-\pi V)},\label{14}
\end{equation}%
from which the specific heat at constant volume can be obtained as
 \begin{equation}
C_V=T\left(\frac{\partial S}{\partial T}\right)_{V}=\frac{S(4S^2-\pi V)}{\pi V-12S^2}.\label{15}
\end{equation}%
It can be seen clearly that the specific heat $C_V\neq0$ since it shares the same factor in its numerator with the denominator of the temperature. So the heat flow along the isochores does not equal to zero. And the isochores are not adiabatic. This case is quite different from those heat engines discussed in the former literatures \cite{Johnson1, Johnson2, Johnson3, Johnson4} whose isochores and adiabats are identical. If one just follows the procedure in the former literatures to calculate the efficiency and ignore this subtle difference, he will certainly make a mistake. Detailed discussions on this issue will be presented in the Discussion Section.

On the other hand, if one retains the standard definition of the thermodynamic volume by introducing a new thermodynamic parameter $R=r_0$ associated with the renormalization length scale as reviewed in Sec.\ref{Sec2}, the situation is much simpler. From Eqs. (\ref{4}) and (\ref{11}), one can soon draw the conclusion that $C_V=0$ which makes the isochores and adiabats identical. Utilizing Eqs. (\ref{3}) and (\ref{4}), the specific heat at constant pressure can be obtained as
 \begin{equation}
C_P=T\left(\frac{\partial S}{\partial T}\right)_{P}=\frac{\pi T}{16P}\left(1+\frac{\sqrt {\pi}T}{\sqrt{\pi T^2+
2PQ^2}}\right).\label{16}
\end{equation}%
By integrating the expression of $C_P$ from $T_1$ to $T_2$, one can obtain the heat input
\begin{eqnarray}
Q_H&=&\int^{T_2}_{T_1}C_PdT=\frac{Q^2}{16}\log\left(\frac{\sqrt {\pi}T_1+\sqrt{\pi T_1^2+2P_1Q^2}}{\sqrt {\pi}T_2+\sqrt{\pi T_2^2+2P_1Q^2}}\right) \nonumber
\\
&+&\frac{\pi (T_2^2-T_1^2)+\sqrt{\pi}(T_2\sqrt{2P_1Q^2+\pi T_2^2}-T_1\sqrt{2P_1Q^2+\pi T_1^2})}{32P_1} .\label{17}
\end{eqnarray}%
The work along the cycle can be calculated as
 \begin{equation}
W=(V_2-V_1)(P_1-P_4)=\frac{4}{\pi}(P_1-P_4)(S_2^2-S_1^2),\label{18}
\end{equation}%
where $S$ is the function of $T$ and $P$. From Eqs. (\ref{3}) and (\ref{4}), $S$ can be expressed as
 \begin{equation}
S=\frac{\pi T+\sqrt{\pi}\sqrt{2PQ^2+\pi T^2}}{16P}.\label{19}
\end{equation}%
Unlike the cases in former literature, the analytic form of $S(T,P)$ can be obtained here, enabling us to fortunately derive the exact expression of heat engine efficiency for charged BTZ black holes. That is
 \begin{equation}
\eta=\frac{W}{Q_H}=(1-\frac{P_4}{P_1}) \times \left[1+\frac{2P_1Q^2\log\left(\frac{\sqrt {\pi}T_2+\sqrt{\pi T_2^2+2P_1Q^2}}{\sqrt {\pi}T_1+\sqrt{\pi T_1^2+2P_1Q^2}}\right)}{B(T_1,T_2,P_1)} \right],\label{20}
\end{equation}%
where
 \begin{eqnarray}
B(T_1,T_2,P_1)&=&\pi (T_2^2-T_1^2)+\sqrt{\pi}\left(T_2\sqrt{2P_1Q^2+\pi T_2^2}-T_1\sqrt{2P_1Q^2+\pi T_1^2}\right) \nonumber
\\
&-&2P_1Q^2\log\left(\frac{\sqrt {\pi}T_2+\sqrt{\pi T_2^2+2P_1Q^2}}{\sqrt {\pi}T_1+\sqrt{\pi T_1^2+2P_1Q^2}}\right).\label{21}
\end{eqnarray}%

Note that in the above calculation, we have follow the scheme where $(T_1, T_2, P_1, P_4)$ is specified as operating parameters in the heat engine cycle. Similarly, if one follow another scheme where $(T_2, T_4, V_2, V_4)$ is specified as operating parameters, one can derive the corresponding input heat $Q_H$, the work $W$ and the efficiency $\eta$ as
  \begin{eqnarray}
Q_H&=&\frac{(Q^2+8\sqrt{\pi V_2}T_2)(V_2-V_4)}{32V_2}-\frac{Q^2}{32}\log\left(\frac{V_2}{V_4}\right),\label{22} \\
W&=&-\frac{Q^2(V_2-V_4)^2}{32V_2V_4}+\frac{\sqrt{\pi}(T_2\sqrt{V_4}-T_4\sqrt{V_2})(V_2-V_4)}{4\sqrt{V_2V_4}},\label{23} \\
\eta&=&(1-\frac{V_4}{V_2})\times\frac{V_2[Q^2(V_2-V_4)+8\sqrt{\pi V_2V_4}(T_4\sqrt{V_2}-T_2\sqrt{V_4})]}{Q^2V_2V_4\log\left(\frac{V_2}{V_4}\right)-V_4(V_2-V_4)(Q^2+8\sqrt{\pi V_2}T_2)}.\label{24}
\end{eqnarray}%

To examine whether the results for the above two schemes are correct, one can double check with the exact formula proposed in Ref. \cite{Johnson4}, which reads
 \begin{equation}
\eta_{f}=1-\frac{M_3-M_4}{M_2-M_1},\label{25}
\end{equation}%
where $M_1, M_2, M_3, M_4$ denote the mass of the black hole evaluated at the four corners of the cycle respectively and we use the subscript "$f$" to denote the heat engine efficiency derived from the exact formula.
Substituting Eq. (\ref{10}) into Eq. (\ref{25}) and utilizing Eq. (\ref{4}), one can obtain
\begin{equation}
\eta_{f}=\frac{64(P_1-P_4)(S_1-S_2)(S_1+S_2)}{64P_1(S_1-S_2)(S_1+S_2)+\pi Q^2\log\left(\frac{S_2}{S_1}\right)},\label{26}
\end{equation}%
For both the schemes that $(T_1, T_2, P_1, P_4)$ and $(T_2, T_4, V_2, V_4)$ are specified as operating parameters respectively, it can be examined that the results of Eq. (\ref{26}) which utilize the exact efficiency formula agree with Eqs. (\ref{20}) and (\ref{24}).

Moreover, we are curious about how the efficiency of our heat engine varies from the rectangle cycle considered here to Carnot cycle. The well-known Carnot cycle consists of two isotherm and two adiabats. The engine expands along an isotherm and an adiabat, then contracts along an isotherm and uses an adiabat to close the path. The Carnot efficiency reads
 \begin{equation}
\eta_C=1-\frac{T_C}{T_H},\label{29}
\end{equation}%
where $T_H, T_C$ denote the temperatures for the two isotherms respectively.
For our heat engine, one can choose $T_H=T_2, T_C=T4$. Then the leading term in Eq. (\ref{20}) can be derived as
 \begin{equation}
1-\frac{P_4}{P_1}=1-\frac{V_2}{V_4}\times\frac{8\sqrt{\pi V_4}T_C+Q^2}{8\sqrt{\pi V_2}T_H+Q^2} .\label{30}
\end{equation}%
When $Q=0$, or the term $8\sqrt{\pi V}T$ is large enough that the $Q^2$ term can be omitted, Eq. (\ref{30}) approaches
 \begin{equation}
1-\frac{P_4}{P_1}\rightarrow1-\frac{T_C}{T_H}\left(\frac{V_2}{V_4}\right)^{1/2},\label{31}
\end{equation}%
which extends the result of Ref. \cite{Belhaj} to three-dimensional spacetime.

\section{Discussions}
\label{Sec4}
In this section, we would like to have a few comments on the two competing thermodynamic approaches introduced in ~\cite{Frassino2} from the perspective of the heat engine efficiency. 

As shown in Sec.\ref{Sec3}, the heat engine efficiency of charged BTZ black holes exactly matches the general formula proposed in Ref. \cite{Johnson4} if one retains the standard definition of the thermodynamic volume by introducing a new thermodynamic parameter $R=r_0$ associated with the renormalization length scale. However, the situation will be more complicated if we insist that $dM=TdS+VdP+\Phi dQ$ and define the relevant thermodynamic quantities as Eqs. (\ref{6})-(\ref{8}). Some qualitative discussion will be presented below.

From Eqs. (\ref{4}) and (\ref{6}), one can easily obtain $4S^2>\pi V$. So Eq. (\ref{15}) suggests that $C_V<0$, implying that the isochores are not adiabatic.

Utilizing Eqs. (\ref{4}), (\ref{6}) and (\ref{14}), one can conclude that the temperature $T$ increases with $P$ when $V$ is fixed. Then we can draw the conclusion that $T_1>T_4, T_2>T_3$ for the cycle considered in Fig.\ref{fg1}. So the heat engine absorbs heat during the process $2\rightarrow3$ and releases heat during the process $4\rightarrow1$. And the efficiency should be calculated as
 \begin{equation}
\eta'=\frac{W'}{Q_{1\rightarrow2}+Q_{2\rightarrow3}}.\label{32}
\end{equation}%

Utilizing Eqs. (\ref{4}) and (\ref{6}), the work $W'$ can be obtained as
 \begin{eqnarray}
W'&=&(V_2-V_1)(P_1-P_4)=\left[\left(\frac{4S_2^2}{\pi}-\frac{Q^2}{32P_1}\right)-\left(\frac{4S_1^2}{\pi}-\frac{Q^2}{32P_1}\right)\right]\times(P_1-P_4) \nonumber
\\
&=&\frac{4}{\pi}(P_1-P_4)(S_2^2-S_1^2),\label{33}
\end{eqnarray}%

Since the definitions of the temperature, the thermodynamic pressure and the entropy are the same for both thermodynamic approaches, Eqs. (\ref{16}), (\ref{17}) and (\ref{19}) also hold for the case here. Then we have
 \begin{equation}
Q_{1\rightarrow2}=Q_H, \;\;\;\; W'=W\label{34}
\end{equation}%

The heat engine efficiency can also be derived from the exact formula \cite{Johnson4} as
 \begin{equation}
\eta'_{f}=1-\frac{M_3-M_4}{M_2-M_1},\label{35}
\end{equation}%
where $M$ should be defined as Eq. (\ref{8}) instead of Eq. (\ref{10}). And Eq. (\ref{8}) can be rewritten as
 \begin{equation}
M=\frac{4PS^2}{\pi}-\frac{Q^2}{32}\left[\log \left(\frac{32P}{\pi}\right)+2\log S\right],\label{36}
\end{equation}%
Substituting Eq. (\ref{36}) into Eq. (\ref{35}), one can show that
 \begin{equation}
\eta'_{f}=\frac{64(P_1-P_4)(S_1-S_2)(S_1+S_2)}{64P_1(S_1-S_2)(S_1+S_2)+\pi Q^2\log\left(\frac{S_2}{S_1}\right)}=\eta_{f}.\label{37}
\end{equation}%
From Eqs. (\ref{32}), (\ref{34}) and (\ref{37}), it is not difficult to deduce that
 \begin{equation}
\eta'<\eta'_{f}.\label{38}
\end{equation}%
This result suggests that the thermodynamic approach with Eqs. (\ref{6})-(\ref{8}) does not agree with the general formula proposed in Ref. \cite{Johnson4}. From this point of view, the heat engine efficiency discussed in this paper may serve as a criterion to discriminate the two thermodynamic approaches introduced in ~\cite{Frassino2} and our result seems to support the approach which introduces a new thermodynamic parameter $R=r_0$.

\section{Conclusions}
\label{Sec5}
    We define a new kind of heat engine via three-dimensional charged BTZ black holes. Specifically, we consider a rectangle cycle in the $P-V$ plane and investigate the efficiency of the heat engine from two different perspectives. As shown in this paper, the three-dimensional charged BTZ black hole spacetime is quite subtle and needs to be more careful. This situation occurs if we insist that $dM=TdS+VdP+\Phi dQ$ and define the corresponding thermodynamic volume. Along the isochores the heat flow does not equal to zero since the specific heat $C_V\neq0$. This point differs from those heat engines discussed in the former literatures \cite{Johnson1, Johnson2, Johnson3, Johnson4} whose isochores and adiabats are identical. So one cannot simply follow the procedure in the former literatures to calculate the efficiency.

    On the other hand, if one introduces a new thermodynamic parameter $R=r_0$ associated with the renormalization length scale, one can retain the standard definition of the thermodynamic volume and the isochores and adiabats become identical. We follow two schemes that $(T_1, T_2, P_1, P_4)$ and $(T_2, T_4, V_2, V_4)$ are specified as operating parameters respectively in the heat engine cycle. We obtain the analytic expression of the three-dimensional charged BTZ black hole heat engine for both schemes. Moreover, we double check with the exact formula proposed in Ref. \cite{Johnson4}. It is shown that the results we obtain for the two schemes are in accordance with those utilizing the exact efficiency formula, thus providing the first specific example for the sound correctness of the exact efficiency formula. We argue that the three-dimensional charged BTZ black hole can be viewed as a toy model for further investigation of the properties of heat engine. Furthermore, we compare our result with that of the Carnot cycle and extend the result of Ref. \cite{Belhaj} to three-dimensional spacetime. In this sense, the result in this paper would be complementary to those obtained in four-dimensional spacetime \cite{Johnson1} or ever higher \cite{Belhaj}.

    Last but not the least, the heat engine efficiency discussed in this paper may serve as a criterion to discriminate the two thermodynamic approaches introduced in Ref.~\cite{Frassino2} and our result seems to support the approach which introduces a new thermodynamic parameter $R=r_0$.

\acknowledgments We would like to express our sincere gratitude to the referee whose deep physical insight has improved the quality of this manuscript significantly. This research is supported by National Natural Science Foundation of China (Grant No.11605082), and in part supported by Natural Science Foundation of Guangdong Province, China (Grant Nos.2016A030310363, 2016A030307051, 2015A030313789).


\begin{thebibliography}{99}

\bibitem{Caldarelli}
 M. M. Caldarelli, G. Cognola and D. Klemm, Thermodynamics of Kerr-Newman-AdS Black Holes and Conformal Field Theories, Class. Quant. Grav. 17 (2000)399-420

\bibitem{Kastor}
 D. Kastor, S. Ray, and J. Traschen, Enthalpy and the Mechanics of AdS Black Holes, Class. Quant. Grav. 26 (2009) 195011

\bibitem{Dolan1}
  B. Dolan, The cosmological constant and the black hole equation of state, Class. Quant. Grav. 28 (2011) 125020

\bibitem{Dolan2}
B. P. Dolan, Pressure and volume in the first law of black hole thermodynamics, Class. Quant. Grav. 28 (2011) 235017

\bibitem{Dolan3}
 B. P. Dolan, Compressibility of rotating black holes, Phys. Rev. D 84 (2011) 127503

 \bibitem{Cvetic}
M. Cvetic, G. Gibbons, D. Kubiznak, and C. Pope, Black Hole Enthalpy and an Entropy Inequality for the Thermodynamic Volume, Phys. Rev. D 84 (2011) 024037



\bibitem{Gunasekaran}
S. Gunasekaran, R. B. Mann and D. Kubiznak, Extended phase space thermodynamics for charged and rotating black holes and Born-Infeld vacuum polarization, JHEP 1211(2012)110

\bibitem{Kubiznak}
D. Kubiz\v{n}\'{a}k and  R. B. Mann, $P$-$V$ criticality of charged AdS black holes, JHEP 1207(2012)033

\bibitem{Frassino}
A. M. Frassino, D. Kubiz\v{n}\'{a}k, R. B. Mann and F. Simovic, Multiple Reentrant Phase Transitions and Triple Points in Lovelock Thermodynamics, JHEP 1409 (2014) 080

\bibitem{Hennigar}
R. A. Hennigar and R. B. Mann, Reentrant phase transitions and van der Waals behaviour for hairy black holes, Entropy 17(2015)8056-8072

\bibitem{Altamirano2}
 N. Altamirano, D. Kubiz\v{n}\'{a}k, R. Mann and Z. Sherkatghanad, Kerr-AdS analogue of tricritical point and solid/liquid/gas phase transition, Class. Quant. Grav. 31(2014) 042001

 \bibitem{Shaowen1}
S. W. Wei, Y. X. Liu, Triple points and phase diagrams in the extended phase space of charged Gauss-Bonnet black holes in AdS space, Phys. Rev. D 90 (2014)044057

\bibitem{Kubiznak2}
D. Kubiz\v{n}\'{a}k, R. B. Mann and M. Teo, Black hole chemistry: thermodynamics with Lambda, arXiv:1608.06147

\bibitem{Johnson1}
C. V. Johnson, Holographic Heat Engines, Class. Quant. Grav. 31(2014)205002

\bibitem{Karch}
A. Karch and B. Robinson, Holographic Black Hole Chemistry, JHEP 12(2015)073

\bibitem{Johnson2}
C. V. Johnson, Gauss-Bonnet Black Holes and Holographic Heat Engines Beyond Large N, Class. Quant. Grav. 33 (2016) 215009

\bibitem{Johnson3}
C. V. Johnson, Born-Infeld AdS Black Holes as Heat Engines, Class. Quant. Grav. 33 (2016) 135001

\bibitem{Sadeghi1}
J.~Sadeghi and Kh.~Jafarzade, Heat Engine of black holes, arXiv:1504.07744

\bibitem{Belhaj}
A. Belhaj, M. Chabab, H. El~Moumni, K. Masmar, M. B. Sedra, and A. Segui, On Heat Properties of AdS Black Holes in Higher Dimensions, JHEP 05(2015)149

\bibitem{Setare}
M.~R. Setare and H.~Adami, Polytropic black hole as a heat engine, Gen. Rel. Grav. 47 (2015)133



\bibitem{Sadeghi2}
J.~Sadeghi and Kh.~Jafarzade, The correction of Ho\v{r}ava-Lifshitz black hole from holographic engine, arXiv:1604.02973

\bibitem{Bhamidipati}
C. Bhamidipati and P. K. Yerra, Heat Engines for Dilatonic Born-Infeld Black Holes, arXiv:1606.03223

\bibitem{Caceres}
E. Caceres, P. H. Nguyen and J. F. Pedraza, Holographic entanglement entropy and the extended phase structure of STU black holes, JHEP 1509 (2015) 184

\bibitem{Johnson4}
C. V. Johnson, An Exact Efficiency Formula for Holographic Heat Engines, Entropy 18 (2016) 120

\bibitem{weishaowen}
S. W. Wei and Y. X. Liu, Implementing black hole as efficient power plant, arXiv:1605.04629


\bibitem{Zanelli}
M. Ba\~{n}ados, C. Teitelboim and J. Zanelli, The black hole in three dimensional space time, Phys. Rev. Lett. 69 (1992) 1849-1851



\bibitem{Zanelli2}
C. Martinez, C. Teitelboim and J. Zanelli, Charged rotating black hole in three spacetime dimensions, 	Phys. Rev. D 61 (2000) 104013


\bibitem{carr}
B. Carr, J. Mureika and P. Nicolini, Sub-Planckian Black holes and the generalized uncertainty principle, JHEP 1507 (2015) 052

\bibitem{Frassino2}
A. M. Frassino, R. B. Mann and J. R. Mureika, Lower-dimensional black hole chemistry, Phys. Rev. D 92 (2015)124069


\bibitem{Grumiller}
D. Grumiller, R. McNees and J. Salzer, Cosmological constant as confining $U(1)$ charge in two-dimensional dilaton gravity, Phys. Rev. D 90 (2014)044032


\bibitem{Cadoni}
M. Cadoni, M. Melis and M. R. Setare, Microscopic entropy of the charged BTZ black hole, Class. Quant. Grav. 25(2008)195022






\end{thebibliography}
\end{document}